# Lightweight Cluster-Based Federated Learning for Intrusion Detection in Heterogeneous IoT Networks

Saadat Izadi, Mahmood Ahmadi*

Computer Engineering and Information Technology Department, Razi Univeristy, Kermanshah, Iran

Email: s.izadi@razi.ac.ir, m.ahmadi@razi.ac.ir

*Corresponding author

**Abstract**

The rise of heterogeneous Internet of Things (IoT) devices has raised security concerns due to their vulnerability to cyberattacks. Intrusion Detection Systems (IDS) are crucial in addressing these threats. Federated Learning (FL) offers a privacy-preserving solution, but IoT heterogeneity and limited computational resources cause increased latency and reduced performance. This paper introduces a novel approach Cluster-based federated intrusion detection with lightweight networks for heterogeneous IoT designed to address these limitations. The proposed framework utilizes a hierarchical IoT architecture that encompasses edge, fog, and cloud layers. Intrusion detection clients operate at the fog layer, leveraging federated learning to enhance data privacy and distributed processing efficiency. To enhance efficiency, the method employs the lightweight MobileNet model alongside a hybrid loss function that integrates Gumbel-SoftMax and SoftMax, optimizing resource consumption while maintaining high detection accuracy. A key feature of this approach is clustering IoT devices based on hardware similarities, enabling more efficient model training and aggregation tailored to each cluster's computational capacity. This strategy not only simplifies the complexity of managing heterogeneous data and devices but also improves scalability and overall system performance. To validate the effectiveness of the proposed method, extensive experiments were conducted using the ToN-IoT and CICDDoS2019 datasets. Results demonstrate that the proposed approach reduces end-to-end training time by 2.47× compared to traditional FL methods, achieves 2.16× lower testing latency, and maintains exceptionally high detection accuracy of 99.22% and 99.02% on the ToN-IoT and CICDDoS2019 datasets, respectively.

## 1. Introduction

The rapid expansion of the Internet of Things (IoT) has led to an unprecedented surge in connected devices, with projections estimating 500 billion IoT devices by 2030, including 90% of vehicles being part of IoT networks [1]. While this growth brings convenience and innovation, it also introduces serious security risks, as IoT devices remain highly vulnerable to evolving cyber threats [2, 3]. Protecting these devices is crucial for ensuring the safety and functionality of IoT ecosystems. This necessitates the development of robust security mechanisms to counter emerging threats. Intrusion Detection System (IDS) has long been a key defense strategy against IoT security breaches [4,5]. Traditionally, IDSs analyze network traffic or system behavior to detect malicious activities. However, recent advancements have introduced Federated Learning (FL) as a privacy-preserving approach for intrusion detection, enabling distributed IoT devices to collaboratively train a model without sharing raw data [6, 7]. In FL, each device locally updates a model and transmits only the model parameters to a central entity, which aggregates updates to improve overall accuracy [8]. This approach reduces communication overhead, enhances privacy, and provides a scalable solution for attack detection [9, 10].

Despite its advantages, FL-based IDS faces a major challengedue to the heterogeneity in IoT environments [11]. IoT networks consist of diverse devices with varying computational capabilities, energy constraints, and network conditions, which complicates the train of a cohesive intrusion detection model. Furthermore, current FL frameworks assume equal participation from all IoT devices, a notion that is impractical given the disparities in hardware and connectivity [12, 13]. These inconsistencies lead to inefficient training, increased latency, and reduced performance of intrusion detection models. Consequently, an effective FL-based IDS must be capable of adapting to IoT heterogeneity while ensuring high level of security, efficiency, and scalability. The heterogeneity of IoT devices presents additional challenges for intrusion detection, requiring frequent model updates to maintain optimal performance [14].

In particular, data heterogeneity can significantly degrade detection accuracy, making it difficult to differentiate benign from malicious activities [9]. Additionally, variations among devices affect training speed, where some devices complete local training within milliseconds, while others take considerably longer. This imbalance slows down the federated learning (FL) process, leading to latency in attack detection, higher false alarm rates, and an overall reduction in accuracy [15].

In order to effectively address the challenges associated with device and data heterogeneity, federated learning (FL) models must implement mechanisms that facilitate their management. A promising solution is the integration of lightweight neural networks, which offer high efficiency with minimal computational overhead [16, 17]. Unlike deep neural networks, which are often impractical for real-time IoT applications, lightweight architectures are designed for low-latency and energy-efficient processing, making them well-suited for resource-constrained IoT devices [18]. Recent research has investigated the integration of FL with lightweight networks to improve scalability and efficiency. For example, Ahmed et al. [19] proposed a lightweight mini-batch FL approach that reduces training rounds while maintaining high accuracyand a low false alarm rate. However, their work did not evaluate latency, an essential factor for real-world applications. These findings highlight the potential of lightweight networks to to achieve a balance among accuracy, efficiency, and resource constraints in IoT-based intrusion detection systems.

Given the critical nature of IoT security and the challenges posed by heterogeneous environments, this paper introduces a novel lightweight cluster-based federated learning intrusion detection approach for heterogeneous IoT. This method utilizes three key components: federated learning (FL), lightweight neural networks, and a hybrid loss function (Gumbel-SoftMax + SoftMax) to enhance detection accuracy while concurrently minimizing computational overhead. To effectively manage heterogeneity, IoT devices are organized into clusters based on similarities in hardware, facilitating for efficient model training and aggregation. This clustering strategy optimizes resource allocation, enabling the development of high-performance learning models with reduced energy consumption and enhanced security. By leveraging FL, the system aggregates knowledge from distributed IoT devices without compromising data privacy. The proposed model employs MobileNet, a lightweight convolutional neural network (CNN), which significantly alleviates computational complexity and memory usage. MobileNet's Depthwise Separable Convolutions ensure efficient processing, making it suitable for resource-constrained IoT environments. This architecture not only lowers the computational burden on individual devices but also minimizes overall training time and latency, enabling real-time threat adaptation and response. Additionally, integrating Gumbel-SoftMax and SoftMax in the final layer enhances the reliability of predictions by introducing a probabilistic factor, improving detection accuracy while further reducing latency. The proposed method was evaluated using the ToN-IoT [20] and CICDDoS2019 [21] datasets, demonstrating superior convergence, with detection accuracy of 99.22% (ToN-IoT) and 99.02% (CICDDoS2019). The model also achieved a 2.47× reduction in training time and significantly lower latency, underscoring its effectiveness in securing dynamic IoT environments. The principal contributions of this approach are as follows:

- **Cluster-based federated learning:** The proposed method organizes IoT devices into clusters based on their hardware similarities, thereby optimizing the processes of model training and aggregation. This strategy significantly enhances scalability and efficiency in heterogeneous IoT environments by ensuring by ensuring more effective contributions to the learning process from devices with similar capabilities.
- **Lightweight model with hybrid loss function**: By incorporating MobileNet, a lightweight CNN architecture, the system reduces computational overhead, making it suitable for resource-constrained IoT devices. Furthermore, the implementation of a hybrid loss function, combining Gumbel-SoftMax and SoftMax, enhances generalization and mitigates false-positive rates, thereby improving the reliability of intrusion detection mechanisms.
- **Low latency and high accuracy**: The proposed approach significantly reduces latency, enabling faster threat detection and response. It also sustains a high level of detection accuracy, ensuring robust security performance in dynamic IoT environments.

The rest of this paper is organized as follows. Section 2 provides an overview of the existing literature on FL in the context of IoT security. Section 3 elaborates on the methodology of our proposed approach. Section 4 outlines the experimental results, comparing our proposed federated learning approach with established techniques. Section 5 concludes the paper.

## 2. Related works

In recent years, research interest in developing federated learning (FL) based intrusion detection systems (IDS) for IoT networks have grown substantially, owing to the advantages of FL in distributed and privacy-preserving environments. Several significant studies in this domain are summarized below. The study in [22] proposes a stacked, unsupervised federated learning (FL) method for flow-based network intrusion detection within a cross-silo setting. The architecture integrates a deep autoencoder with an energy-based flow classifier in an ensemble configuration. Experimental results show that this approach surpasses both conventional local models and simple cross-evaluation techniques. Moreover, it performs robustly under non-IID data distributions, making it suitable for deployment across heterogeneous network silos. Mothukuri et al. [23] proposed another FL-based system combining Random Forest (RF) and GRUs to detect and classify attacks in IoT environments. Integrating RF with GRUs within the FL framework improved detection performance by reducing prediction errors and lowering false-alarm rates compared to centralized learning methods.

Shukla et al. [24] introduced a cloud-assisted FL framework that incorporates heterogeneous model types for IoT malware detection. Their evaluation on Raspberry Pi devices demonstrated that the proposed design can improve detection accuracy while maintaining relatively low computational overhead. Similarly, Li et al. [25] presented DeepFed, an FL scheme for cyber-physical systems that leverages the complementary strengths of CNNs and GRUs. Their results demonstrated improved intrusion detection performance over existing FL-based frameworks. A comprehensive survey in [26] evaluated multiple datasets to benchmark FL techniques applied to IoT security. To address persistent limitations in existing FL-based IDS solutions, the work in [27] proposed DEAFL-ID, an optimized system for device selection, training, and detection in heterogeneous Industrial IoT (IIoT) environments. Simulation results confirmed that DEAFL-ID significantly reduces training costs while achieving superior detection performance compared to previous methods.

Despite these advancements, a major limitation of FL remains its assumption of identical on-device models a constraint that becomes problematic in heterogeneous IoT systems. To mitigate this, a cloud-based scalable FL method is presented in [28], which was evaluated on Raspberry Pi testbeds. The findings indicate that the system improves malware detection with minimal computational overhead. Furthermore, [11] introduced HetIoT-CNN IDS, a deep learning–based network-traffic IDS tailored for heterogeneous IoT environments, capable of detecting diverse types of attacks efficiently. Building on these developments, the proposed cluster-based federated intrusion detection system with a lightweight network addresses the dual challenge of data heterogeneity and device-level computational disparities. By combining MobileNet-based lightweight architectures with a federated learning framework and a hybrid loss function (Gumbel-SoftMax + SoftMax), the system achieves high detection accuracy and low false-positive rates while reducing computational complexity and latency.

## 3. Proposed method

This section presents the proposed cluster-based federated intrusion detection system, which integrates federated learning, lightweight deep learning models, and a hierarchical training structure to optimize intrusion detection in heterogeneous IoT environments. The comprehensive architecture of the proposed method is illustrated in Figure 1, while the mathematical formulations underpinning the approach are detailed in Table 1, which provides a comprehensive list of notations along with their corresponding descriptions.

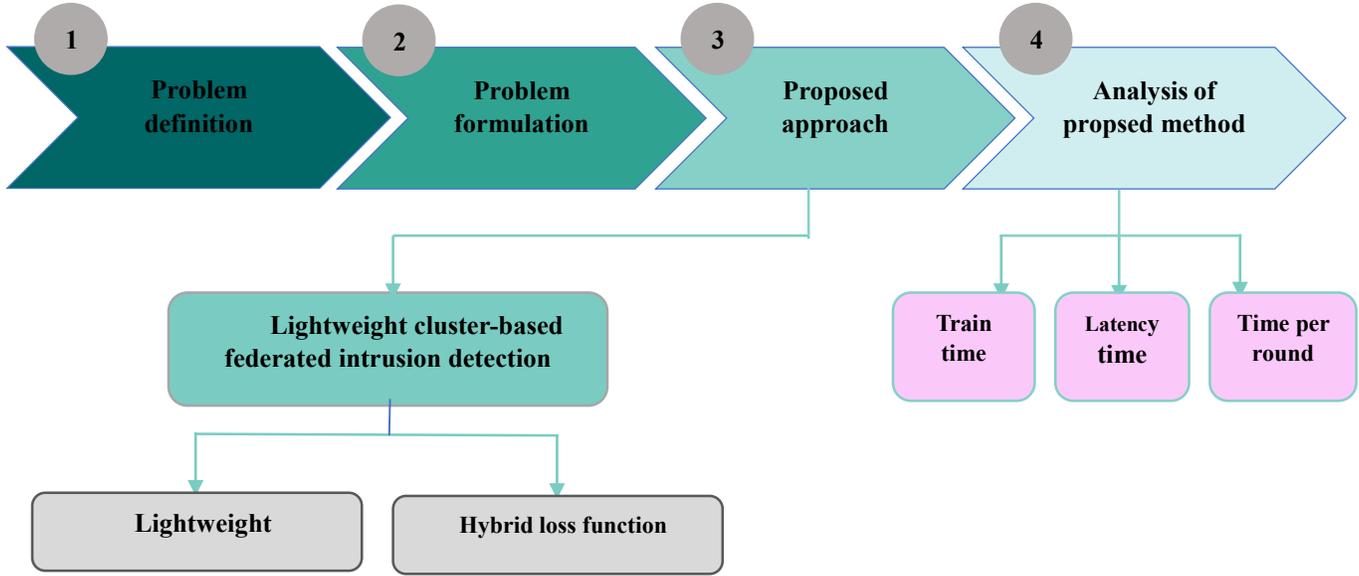

**Figure 1.** Different steps of the developing of the intrusion detection with federated learning in cluster-based environments.

**Table 1**. List of commonly used notations in the proposed method

| Notation | Description |
|---|---|
| $w_{ic}(t)$ | Model weights of device i at round t. |
| $w_c(t)$ | Aggregated model weights of cluster c at round t. |
| $w(t)$ | Global model weights at round t. |
| $\eta$ | Learning rate used in model optimization. |
| $D_{ic}$ | Local dataset of device i in cluster $C_k$. |
| $D_c$ | Total dataset size in cluster $C_k$. |
| D | Total dataset size across all devices. |
| N | Number of IoT devices participating in federated learning. |
| C | Number of clusters in the federated learning model. |
| x | Input data for convolutional layers. |
| W | Weights in the standard convolutional layer. |
| $W_{depth}$, $b_{depth}$ | Weights and biases associated with the depthwise convolution layer. |
| $W_{point}$, $b_{point}$ | Weights and biases for the pointwise convolution layer. |
| F | Number of filters in convolutional layers. |
| $\sigma$ | Activation function used for non-linearity in deep learning. |
| Bandwidth | Network bandwidth between IoT devices and the server. |

### 3.1. System architecture and problem definition

The increasing heterogeneity of IoT devices has led to significant security challenges, particularly in detecting intrusions in resource-constrained environments. Traditional intrusion detection systems (IDS) struggle due to limited device capabilities, data heterogeneity, and high communication overhead. Federated learning (FL) offers a promising solution, but existing approaches do not efficiently address hardware disparities among IoT devices. Therefore, there is a need for a scalable and efficient FL-based IDS that minimizes latency, computational overhead, and false detection rates while ensuring privacy-preserving learning. Figure 2 provides a visual representation of the proposed method. It illustrates the hierarchical structure of the Internet of Things (IoT), consisting of the edge, fog, and cloud levels. In this architecture, intrusion detection clients are located in the fog layer and utilize federated learning. The hypothesis considered in this paper for the proposed idea to be suitable for simulation in the real world is that considering that it has a heterogeneous IoT, and this heterogeneity includes devices and data. Considering that the data is heterogeneous such as ToN-IoT which was deployed to manage the interconnection between the three layers of IoT: Edge, Fog and Cloud layers. In Edge layer, various data types are gathered from IoT and IIoT sensors. In Fog layer, to show the heterogeneity of IoT devices, c types of clusters are organized in this architecture, each of these clusters has different computing resources in terms of CPU/GPU and RAM, and each cluster consists of a number of clients and central server that is located in Cloud layer. In this paper, IoT devices are grouped into clusters based on hardware similarity to address heterogeneity. This hardware similarity is determined using predefined hardware information, such as processing power, memory capacity, and communication bandwidth. It is assumed that this information is either known beforehand or collected during the initial setup phase. No dynamic Configuration-Aware algorithm is used; instead, devices are statically grouped into clusters with similar hardware capabilities.

In heterogeneous Internet of Things systems, a wide variety of devices with different hardware configurations are extensively exposed to cyberattacks. These devices possess varying computational resources, memory, and energy capacities, which generate heterogeneous data and introduce significant complexity in designing effective IDS. In this paper, IoT devices are grouped into clusters based on hardware similarity to address heterogeneity. This hardware similarity is determined using predefined hardware information, such as processing power, memory capacity, and communication bandwidth. It is assumed that this information is either known beforehand or collected during the initial setup phase. No dynamic configuration-aware algorithm is used; instead, devices are statically grouped into clusters with similar hardware capabilities. This paper proposes a federated learning-based approach that leverages the lightweight MobileNet model to optimize resource consumption and reduce latency. By employing a hybrid loss function, the proposed method aims to enhance detection accuracy and effectively identify various types of attacks in heterogeneous IoT environments. To simplify the complexity, it is assumed that IoT devices are grouped into c clusters based on their hardware type, as detailed in Section 4. Each type of hardware is assigned to a specific group or cluster. In this context, a cluster refers to a group of IoT devices with identical hardware. Devices with similar hardware are placed within the same cluster, and the model training and aggregation processes are conducted separately for each cluster. The following sections detail the problem modeling, the lightweight model with the hybrid loss function, and an analysis of the proposed method.

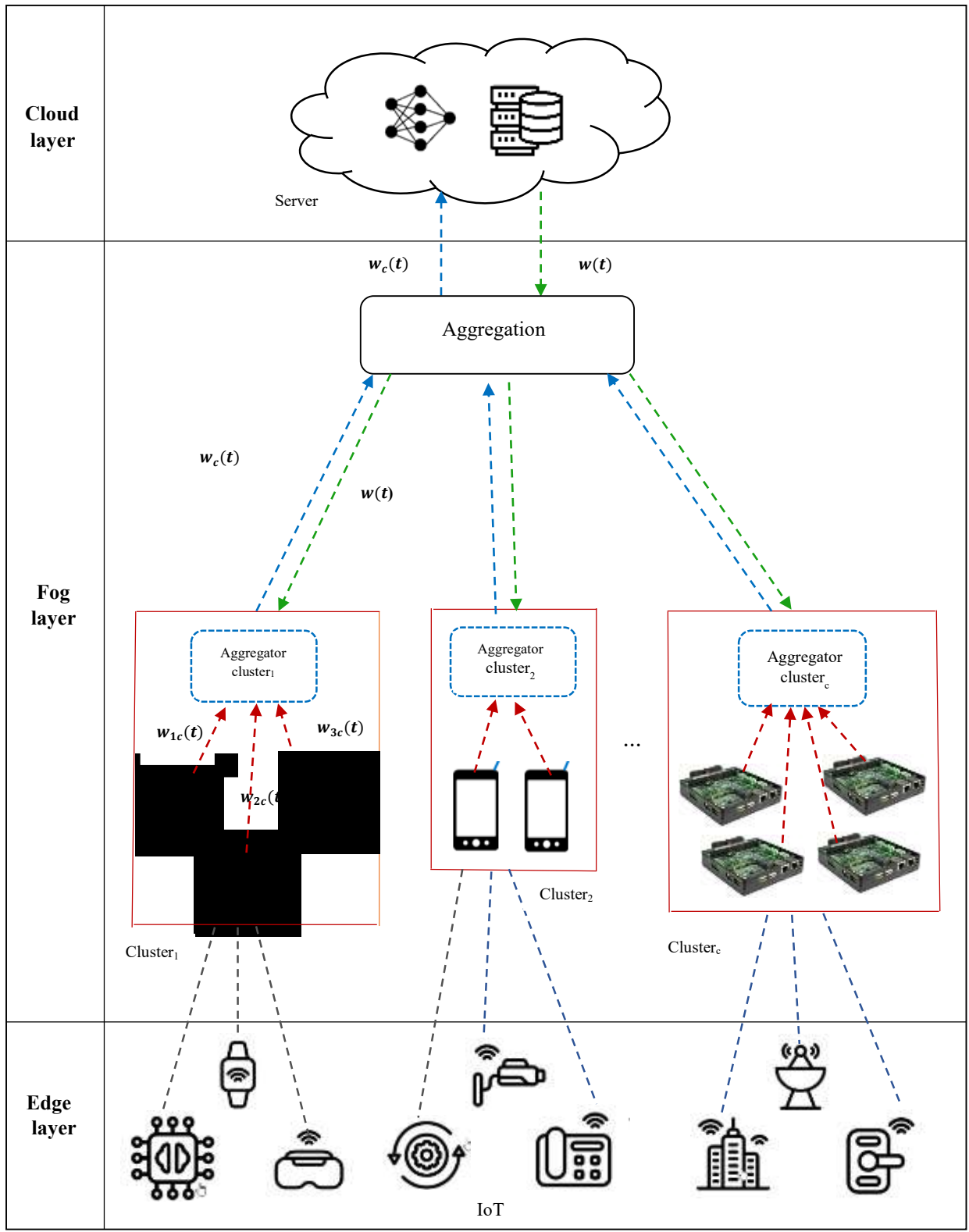

**Figure 2.** Architecture of intrusion detection using federated learning in IoT. The system defines multiple clusters (cluster 1, cluster 2, …, cluster c) each with varying computational resources, distributed across the cloud, fog, and edge layers.

## 3.2. Problem formulation

In this paper, we assume the existence of $C$ clusters, where each cluster comprises $N_c$ devices (clients). The data of each device is represented as $D_{ic}$, where device $i$ belongs to cluster $c$. The aggregated model weights for each cluster are denoted by $w_c$, and the model weights for device $i$ after local training are represented by $w_{ic}$. Initially, data is allocated to different clusters and then further distributed among the devices within each cluster. This distribution is defined as Eq. (1).

$$D = \bigcup_{c=1}^{C} \bigcup_{i=1}^{N_c} \quad (1)$$

During the local training phase, each device $i$ within cluster $c$ updates its model weights using its local data. The weight update formula is as follows Eq. (2):

$$w_{ic}(t+1) = w_{ic}(t) - \eta \nabla L_{ic}(w_{ic}(t); D_{ic}) \quad (2)$$

Where, $w_{ic}(t)$ represents the model weights on device $i$ in cluster $c$ at round $t$, $\eta$ is the learning rate, and $L_{ic}$ denotes the loss function for the data on device $i$ in cluster $c$. After completing local training, the model weights within each cluster are combined to obtain an aggregated model for each cluster. The weight aggregation is performed according to Eq. (3).

$$w_c(t+1) = \sum_{i=1}^{N_c} \frac{|D_{ic}|}{|D_c|} w_{ic}(t+1) \quad (3)$$

where $D_{ic}$ represents the data of device $i$ in cluster $c$, and $D_c$ denotes the total data in cluster $c$. (Eq. (4)).

$$|D_c| = \sum_{i=1}^{N_c} |D_{ic}| \quad (4)$$

In each cluster, a cluster head selected from one of the client devices to coordinate the aggregation process. This node temporarily acts as the head and is responsible for collecting local models from all devices within the cluster, computing the aggregated model using weighted averaging, and forwarding it to the central server. The combined models from each cluster are forwarded to the central server. The central server integrates the combined model weights from the clusters to generate the final model, which is then disseminated to the devices. The weight aggregation between clusters is performed according to Eq.(5).

$$w(t+1) = \sum_{c=1}^{C} \frac{|D_c|}{|D|} w_c(t+1) \quad (5)$$

Where $D$ represents the total data and |D| represents the total number of data samples in the dataset $D$ (Eq. (6)).

$$|D| = \sum_{c=1}^{C} |D_c| \quad (6)$$

The training process is repeated over several rounds to further improve the final model. Each round generally includes the following steps: **local training**: each device i in cluster c trains its model using local data, **intra-cluster aggregation**: the model weights within each cluster are combined to obtain the aggregated model for that cluster, **transferring to central server**: the aggrega1ted models from each cluster are sent to the central server, and **inter-cluster aggregation at central server**: the central server consolidates the aggregated model weights from the clusters to obtain the final model, which is subsequently returned to the devices.

*3.3. Proposed approach : Cluster-based federated learning architecture*

The proposed CB-FLIDS follows a hierarchical approach, organizing IoT devices into clusters based on their computational capabilities (CPU, RAM, power efficiency). This structure consists of three layers: The **edge layer** collects data from IoT and IIoT sensors. Devices are grouped into hardware-aware clusters to ensure efficient training. The **fog layer** serves as an intermediate processing layer, where cluster aggregators perform local model aggregation. It helps reduce communication overhead by handling computations before forwarding model updates to the cloud. The **cloud layer** integrates cluster-level models to generate a global intrusion detection model. Once the global model is created, it is disseminated back to the clusters for continued learning.

*3.3.1 Lightweight MobileNet for efficient intrusion detection*

For IoT intrusion detection with federated learning, reducing latency is one of the primary objectives. To achieve this goal, the use of lightweight networks that can operate with high speed and efficiency in resource-constrained environments is crucial. One such lightweight network is MobileNet. MobileNet is a neural network specifically designed for devices with limited resources, such as IoT devices. By optimizing its architecture, MobileNet enables efficient data processing while minimizing resource consumption. The architecture of MobileNet is engineered to perform data processing more quickly and with lower latency. MobileNet utilizes two primary types of layers: standard convolutional layers and depthwise separable convolutional layers. These layers are alternatively called depthwise convolution and pointwise convolution. The standard convolutional layer in MobileNet is defined as follows (Eq.(7)):

$$\text{Conv}(x) = \sigma(W \cdot x + b) \tag{7}$$

Here, $w$ signifies the weights, and $b$ reperesnts the bias, which help the model extract features from the input data $x$. The activation function $\sigma$ assists the model in learning the nonlinear behavior of the data. The depthwise separable convolutional layer is divided into two parts: depthwise convolution and pointwise convolution. Each of these layers is defined as Eq. (8) and Eq. (9).

$$\text{Conv}(x) = \sigma(\text{DepthwiseConv}(x)) = \sigma((W_{depth} \cdot x + b_{depth}) \cdot W \cdot x + b) \tag{8}$$

$$\text{PointwiseConv}(x) = \sigma(W_{point} \cdot x + b_{point}) \tag{9}$$

where, $W_{depth}$ and $b_{depth}$ represent the weights and biases associated with the depthwise convolution layer, while $W_{point}$ and $b_{point}$ denote the weights and biases for the pointwise convolution layer. The final relationship for the depthwise separable convolution layer is combined into Eq.(10).

$$\text{Conv}(x) = \sigma\left((W_{depth} \cdot x + b_{depth}) \cdot W_{point} + b_{point}\right) \tag{10}$$

In the proposed method, local models are trained on each IoT device using the MobileNet architecture. The weights of these local models are then sent to a central server for aggregation. By utilizing MobileNet, effective and efficient models can be developed that perform well in resource-constrained environments and assist in intrusion detection within IoT networks.

*3.3.2. Hybrid loss function for robust intrusion detection*

To further improve classification performance, we introduce a hybrid loss function, combining SoftMax loss (for attack classification) and Gumbel-SoftMax loss (for enhanced generalization):

$$L_{Hybrid} = \alpha \cdot L_{SoftMax} + (1-\alpha) \cdot L_{Gumbel\text{-}SoftMax} \tag{11}$$

This approach enhances detection accuracy, reduces false positives, and improves resilience against adversarial attacks.

*3.4. Analysis of the proposed method*

To assess the proposed method, several performance indicators including accuracy, recall, and false positive rate were employed. Given that MobileNet has been selected to reduce resource consumption and increase efficiency, this lightweight model is designed to consume fewer computational resources while still providing satisfactory accuracy. In the context of federated learning, particularly for IDS in a heterogeneous Internet of Things (IoT) environment, understanding and calculating key performance parameters are critical for evaluating system efficiency and effectiveness. The following metrics are of primary importance.

*3.4.1. Training Time*

The total training time encompasses the time taken for local training on each client device, the intra-cluster aggregation of model updates, and the communication time to transmit updates to the central server. The overall training time can be summarized as follows:

$$T_{Total}^{Training} = \max\left(T_{Local} + T_{Intra} + T_{Comm}\right) + T_{Server} \tag{12}$$

Which $T_{Local}$ is the local training time on client devices, $T_{Intra}$ is the Intra-cluster aggregation time within clusters, $T_{Comm}$ is the communication time for sending the aggregated model to the central server, and $T_{Server}$ is the server processing time at the central server to aggregate and update the global model.

- **Local training time** or each device $i$ in cluster $c$ (Eq. (13)).

$$\text{Local training time}_i^c = \frac{D_i^c}{R_i^c} \tag{13}$$

Where, $D_i^c$ size of the local dataset on device $i$ in cluster $c$ (in samples), and $R_i^c$ learning rate (training speed) of device $i$ in cluster $c$ (in samples per second).

- **Intra-cluster aggregation time**

This is the time required to aggregate the model weights within the cluster Eq.(14):

$$\text{Intra-cluster aggregation time}^c = N^c_{clients} \cdot T^c_{agg} \qquad (14)$$

Where, $N^c_{clients}$: Number of devices in cluster $c$, and $T^c_{agg}$: Time taken for aggregation in cluster $c$ (in seconds). In federated learning, the intra-cluster aggregation time is determined by the maximum training time among all clients within a cluster, as model aggregation can only occur after all clients have completed their local training. Therefore, the aggregation time for each cluster is calculated as the maximum training time across all clients in that cluster:

$$T^c_{agg} = \max(T_1, T_2, \ldots, T_N) \qquad (15)$$

This approach ensures synchronization across clients, although it may lead to latency due to slower devices.

- **Communication time**

The time taken to communicate the aggregated model to the server Eq.(14):

$$\text{Communication time}^c = \frac{M_c}{B_c} \qquad (16)$$

Where, $M_c$ size of the data sent from cluster $c$ to the central server, and $B_c$ bandwidth of the connection between cluster $c$ and the server. The communication time varies for each client within a cluster due to differences in data sizes. In this study, the communication time is calculated based on the average data size across all clients within a cluster. However, to ensure robust performance analysis, the worst-case communication time (i.e., the client with the maximum data size) is also considered. This approach provides a comprehensive evaluation of latency and ensures that the system can handle variations in data distribution effectively.

Also, the server processing time in federated learning is the time required on the central server to perform two main tasks: aggregation and global model update. Although no direct training occurs on the server, these processes are essential for synchronizing the global model across all clients. The total server processing time can be calculated as follows:

$$Time_{Server} = T_{Agg} + T_{Update} \qquad (17)$$

Where, $T_{Agg}$ is the aggregation time, which represents the time needed to collect and aggregate model updates from all clusters. It is calculated as:

$$T_{Agg} = \frac{D_{Total}}{B_{Server}} \qquad (18)$$

Here, $D_{Total}$ is the total size of data sent from all clusters to the server, and $B_{Server}$ is the bandwidth of the central server. This part of the formula shows that higher bandwidth reduces aggregation time. $T_{Update}$ is the global model update time, which is the time required to update the global model using the aggregated parameters. It is calculated as:

$$T_{Update} = \frac{P}{F_{Server}} \qquad (19)$$

Where P represents the number of parameters in the global model, and $F_{Server}$ is the processing speed of the central server (measured in parameters per second). This indicates that a faster processing speed decreases the model update time. By including $Time_{Server}$ in the overall training time, a more comprehensive estimation of the total training latency is obtained, ensuring that both aggregation and model update delays are accurately represented.

*3.4.2. Testing latency*

Testing latency refers to the time required to process new inputs and return predictions. The total testing latency can be expressed as:

$$\text{Testing latency} = \sum_{c=1}^{C} \left( \text{Inference time}^c + \text{Communication time}^c \right) \quad (20)$$

- **Inference time**

This is the time taken to process new input using the trained model (Eq.(21)) .

$$\text{Inference time}^c = \frac{N_{input}^c}{F^c} \quad (21)$$

Where, $N_{input}^c$ number of input samples for cluster $c$ (in samples), and $F^c$ processing speed of the model in cluster $c$ (in samples per second).

- **Communication time for testing**

The communication time for sending input data to the cluster for processing (Eq.(22 )):

$$\text{Communication time}_c = \frac{M_{test,c}}{B_c} \quad (22)$$

Where. $M_{test,c}$ size of the input data sent to cluster $c$ for processing, and $B_c$ bandwidth of the connection between cluster $c$ and the server.

*3.4.3. Time per round*

Time per round includes the total time taken for one complete cycle of training, which consists of local training, aggregation, and communication.

$$\text{Time per round} = \sum_{c=1}^{C} \left( \text{Local training time}^c + \text{Intra-cluster aggregation time}^c + \text{Transmission time}^c \right) \quad (23)$$

Where, Transmission Time$^c$ is equivalent to the communication time calculated earlier. In the paper assumes a hypothetical bandwidth of 12.5 Mbps for uplink calculations.

## 3. Evaluation result

In the following section, the experimental setup, evaluation metrics, dataset, preprocessing, result and comparison of the proposed method by other related works will be presented.

### 4.1. Experimental setup

The experimental process in this paper was conducted in a computer system with an NVIDIA GPU RTX 3080Ti GPU with an 8GB RAM. The system is powered by an Intel Core i9-7800X processor with a base speed of 3.50 GHz to an ultimate speed of 4.00 GHz, alongside 128 GB DDR4 RAM. The development process employed JupyterLab 3.6.7 IDE and Python programming language, which used commonly employed libraries such as NumPy[29], Pandas[30], and Pytorch [31]. This work employed a simulation context it has 3 cluster nodes, 20 clients and a single server. Besides this, the data distribution patterns to be employed are non-independent and identically distributed (non-IID) in nature. Training is conducted with 80% of the dataset while testing is conducted with 20% of the dataset. MobileNet is employed to conduct the lightweight model. Parameters employed in the FL methods are: local epochs = 50, number of rounds = 10, batch size = 128, learning rate = 0.001, dropout = 0.5, with an Adam optimizer employed. All clients possess diverse computing resources in processor capacity and memory. $T = 0.5$ is employed in this work. Furthermore, $\gamma$ is employed to be equal to 0.6 in continuously distributed labels. Values to be employed in the parameters are explained in Table 3. Client setups in c=3 clusters are explained in Table 2 and 3.

Table 2. Parameter values.

| Parameter | Value |
|---|---|
| Batch normalization | 128 |
| Learning rate | 0.001 |
| Dropout | 0.1 |
| #clients | 10 |
| #Epoch | 50 |
| #Rounds | 10 |
| $\Gamma$ | 0.6 |
| $T$ | 0.5 |
| $\alpha$ | 0.5 |
| c | 3 |

Table 3. Cluster configurations.

| Cluster | Model | CPU/GPU | RAM | Number of Clients |
|---|---|---|---|---|
| Cluster 1 | Raspberry Pi 3 Model B | 1.2GHz | 1GB | 2 |
| Cluster 2 | Raspberry Pi 4 Model B | 1.5GHz | 4GB | 3 |
| Cluster 3 | Raspberry Pi 400 | 1.8GHz | 8GB | 5 |

### 4.2. Performance evaluation metrics

Metrics for measuring the method given are Accuracy, Precision, F1-score, Recall, and TPR. The metrics are derived based on True Positives, True Negatives, False Positives, and False Negatives. The outcomes based on the metrics are discussed in detail in subsequent sections.

- **Accuracy** is a metric specifying the ratio of correct classifications out of total inputs. It is calculated in accordance with Eq. (24).

- **Precision** measure the ratio of elements identified as positive out of total elements labeled as positive. Eq. (25) shows how the measure is calculated as a ratio of accurately classified anomalies and total instances labeled as positive.

- **Recall** is calculated as the ratio of the number of properly classified attacks to the total number of attacks. Recall is computed based on Eq. (26).

- **F1-Score** is a metric for measuring the quality of classifying assaults and is computed as a ratio of properly classified attacks to total expected outcomes for attacks and can be derived based on Eq. (27).

- **True Positive Rate (TPR)** refers to the occurrence when an IDS correctly identifies an action as an attack, and that activity indeed represents an intrusion. TPR, commonly known as the detection rate, can be expressed by Eq. (28).

- **False positive rate (FPR)** refers to the situation when the IDS incorrectly identifies a typical activity as an attack. A false positive can also be referred to as a false alarm rate, which can be computed using Eq. (29).

$$Accuracy = \frac{TP + TN}{TP + TN + FP + FN} \quad (24)$$

$$Precision = \frac{TP}{TP + FP} \quad (25)$$

$$Recall = \frac{TP}{TP + FN} \quad (26)$$

$$F1 - score = 2 * \frac{Recall * precision}{Recall + precision} \quad (27)$$

$$True\ Positive\ Rate\ (FPR) = \frac{TP}{TP+FN} \quad (28)$$

$$False\ Positive\ Rate\ (FPR) = \frac{FP}{FP+TN} \quad (29)$$

*4.3. Dataset*

The methodology was confirmed by adopting the dataset known as the ToN-IoT dataset [20], which is used to acquire and analyze an ample variety of sources of IoT (Internet of Things) and IIoT (Industrial Internet of Things) data. UNSW at ADFA developed an ample network established to acquire the dataset known as ToN-IoT. This physical system included virtual devices, physical systems, cloud infrastructure, and IoT sensors to reflect an assortment of heterogeneous sources of data. This dataset consists of various forms of diverse information gathered through various sources including sensor data gathered through interfaced devices, system log messages in both systems (Linux and Windows systems) and system networking traffic. For testing effectiveness and correctness of various cybersecurity tools developed by AI methods, samples have been gathered from an online network system. This dataset has 43 attributes and an ample amount of regular traffic plus nine types of attacks. 2: As an additional effectiveness evaluation was required to prove this work superior to earlier methods, this work employed dataset CICDDoS2019 [21], collected by the Canadian Institute for Cybersecurity and the University of New Brunswick. This dataset has 86 attributes. This dataset is used in most attack detection-related research in HetIoT systems [32]. Figure 3 and Figure 4 present these datasets in greater detail including attack names and number of samples.

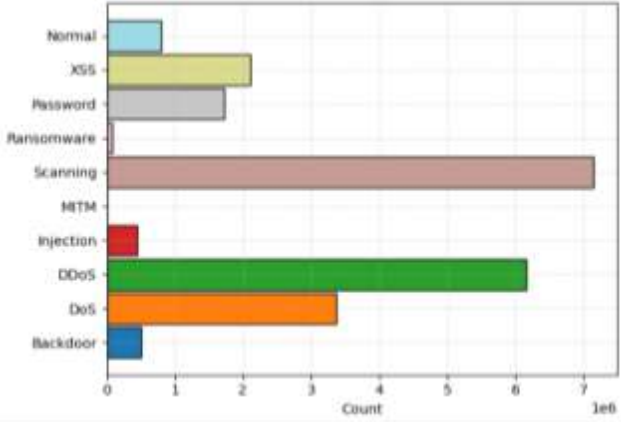
**Figure3**. ToN-IoT

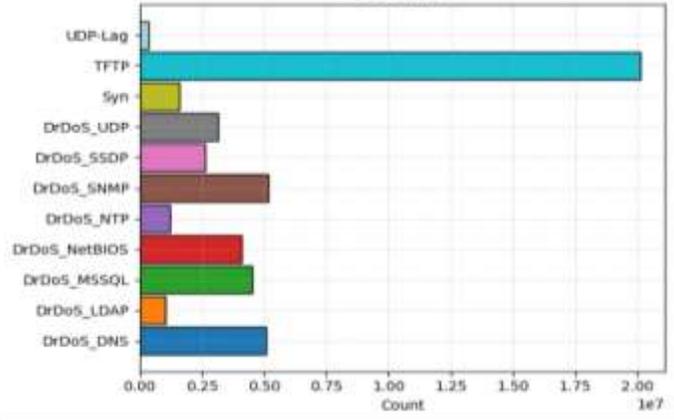
**Figure 4**. CICDDoS2019

*4.4. Preprocessing*

Preprocessing is essential in federated learning to optimize the efficiency and effectiveness of collaborative learning among distributed devices. Due to the decentralized structure of federated learning, where data is kept on local devices and only model updates are sent to a central server, preprocessing tasks are typically carried out locally. This requires thoughtful selection of preprocessing methods that uphold data privacy and security while also guaranteeing the quality and uniformity of the models on all devices involved. Preprocessing tasks like managing missing values, scaling features, converting object to numeric, normalization, and standardization may require adjustments to align with the distributed nature of federated learning. Normalization scales feature values to a range of 0 to 1, while standardization transforms feature values to have a mean of 0 and a standard deviation of 1. These techniques are crucial for data preparation in federated learning settings. The techniques ensure data consistency and enhance model training efficiency across various data sources, while also addressing privacy concerns using methods such as differential privacy. The formula for normalization is:

$$X' = \frac{x - x_{min}}{x_{max} - x_{min}} \quad (30)$$

where x denotes the original value of the feature, $x_{min}$ shows the minimum value of the feature, and $x_{max}$ is the maximum value of the feature. The formula for Z-Score standardization is:

$$Z = \frac{x - \mu}{\sigma} \quad (31)$$

where X is the current feature value, $\mu$ is the mean, and $\sigma$ is the standard deviation.

*4.5. Results*

This section presents the experimental evaluation of the proposed intrusion detection system, designed to enhance security in heterogeneous IoT environments by leveraging federated learning (FL). The system's performance was rigorously tested using two benchmark datasets, ToN-IoT and CICDDoS2019, ensuring a comprehensive assessment across diverse intrusion scenarios. Figure 5 illustrates the evaluation metrics, including accuracy, precision, recall, and F1-score, measured over multiple training rounds. As the graphs depict, the model demonstrates consistent and significant improvement across all evaluation metrics, reflecting its ability to effectively learn from decentralized data sources. Initially, accuracy and precision start at relatively lower values, as the model undergoes its initial learning phase. However, with increasing training rounds, these metrics progressively improve, surpassing 99% by the final iterations. This highlights the system's ability to precisely distinguish between normal and malicious activity, ensuring a high detection rate with minimal false positives, a crucial factor in real-time intrusion detection.

Similarly, the recall metric shows steady enhancement, validating the model's capability to correctly identify actual intrusion attempts while minimizing false negatives. This improvement is essential for reducing undetected security threats that could otherwise compromise an IoT network. The F1-score, which balances precision and recall, also exhibits continuous growth, emphasizing the overall robustness and reliability of the proposed system. These findings affirm that federated learning effectively mitigates the challenges of IoT data heterogeneity, allowing the model to leverage distributed data sources while preserving privacy and maintaining high detection accuracy. Beyond achieving superior detection performance, the results underscore the critical role of lightweight neural networks in accelerating federated learning convergence. The integration of MobileNet, a computationally efficient deep learning model, plays a pivotal role in reducing training overhead, making real-time deployment feasible even in resource-constrained IoT environments. The experimental findings confirm that lightweight networks significantly enhance training speed, ensuring faster convergence without compromising accuracy. Moreover, the incorporation of advanced loss functions, such as Gumbel-SoftMax, further strengthens the model's generalization capability. This technique enhances the system's resilience against evolving cyber threats, enabling it to effectively detect a wide range of attack patterns while minimizing misclassifications. The combination of federated learning, lightweight architectures, and optimized loss functions provides a highly efficient, scalable, and adaptive solution for securing IoT networks against sophisticated cyber threats.

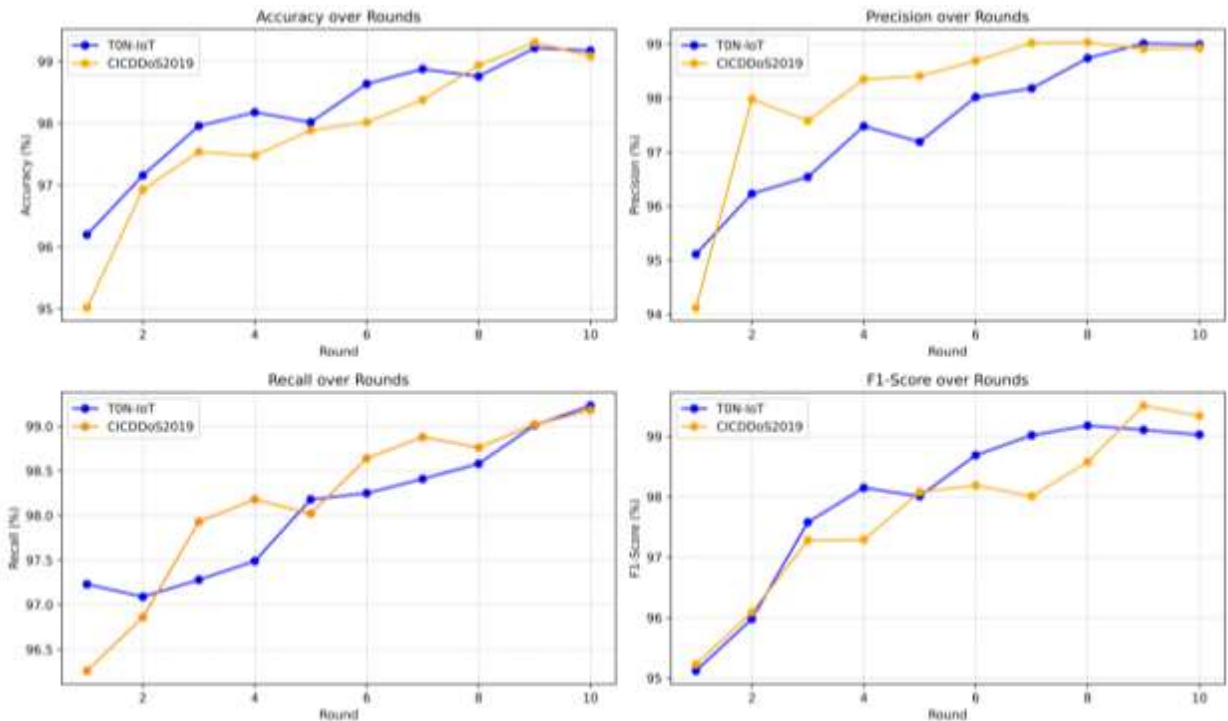

Figure 5: Evaluation metrics of training rounds in proposed method for ToN-IoT and CICDDOs2019

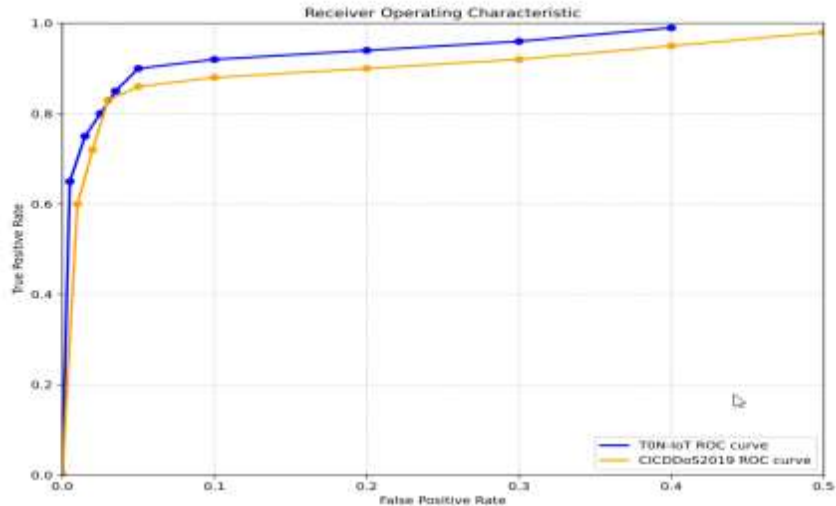

Figure 6. ROC curves of the proposed method for ToN-IoT and CICDDoS2019 datasets.

Figure 6 shows the Receiver Operating Characteristic (ROC) curves for the proposed Cluster-Based Federated Intrusion Detection with Lightweight Networks for Heterogeneous IoT applied to the ToN-IoT and CICDDoS2019 datasets. The ROC curve is a visual representation of the performance of a binary classifier as its decision threshold changes. The relationship between TPR and FPR, visualized in a plot, provides insights into the classification model's performance. The curves illustrate the model's high true positive rate (TPR) and low false positive rate (FPR), suggesting its strong performance in distinguishing between normal and attack classes. The ROC curve for ToN-IoT consistently outperforms CICDDoS2019, highlighting superior detection capabilities in heterogeneous IoT environments. The steep initial rise in the curves demonstrates the model's efficiency in quick and accurate detection. This approach positively impacts the ROC by enhancing the model's ability to achieve a higher true positive rate with a lower false positive rate, thus improving overall detection performance and ensuring reliable and efficient intrusion detection in diverse IoT environments.

Figures 7 and 8 present the confusion matrices for the ToN-IoT and CICDDoS2019 datasets, respectively, providing an in-depth evaluation of the proposed model's classification performance. A strong diagonal presence in both matrices indicate a high detection accuracy, with the majority of attacks correctly identified. In Figure 7, the model demonstrates exceptional performance in detecting large-scale threats such as Scanning and DdoS, which both of which pose significant challenges in IoT environments due to their dynamic attack patterns. Similarly, in Figure 8, the model continues to showcase its robustness, maintaining precise classification across multiple attack types. A particularly noteworthy achievement is the detection of DDoS attacks, which are among the most complex due to their distributed and high-volume nature. In both datasets, the model effectively minimizes misclassification and achieves a high recall rate for DDoS, underscoring its capacity to handle large-scale network intrusions. This success is attributed to the proposed cluster-based federated learning approach, which organizes IoT devices based on hardware similarities, ensuring more precise model aggregation and reducing classification inconsistencies.

Considering that clustering has been used to simulate the heterogeneity of clients. The evaluation results of all three clusters for ToN-IoT and CISDDOS2019 dataset are shown in Figure 9. Figure 9 presents the performance metrics for proposed method across three different cluster configurations for both datasets. For ToN-IoT, cluster 3, with the highest hardware specifications (1.8GHz CPU, 8GB RAM, 5 clients), achieves the best results, showing an accuracy of 99.28%, precision of 99.35%, recall of 99.13%, and an F1-score of 99.55%. This exceptional performance is due to the cluster's powerful computational resources, which facilitate complex computations and rapid data processing, enabling the federated learning model to converge quickly and efficiently.

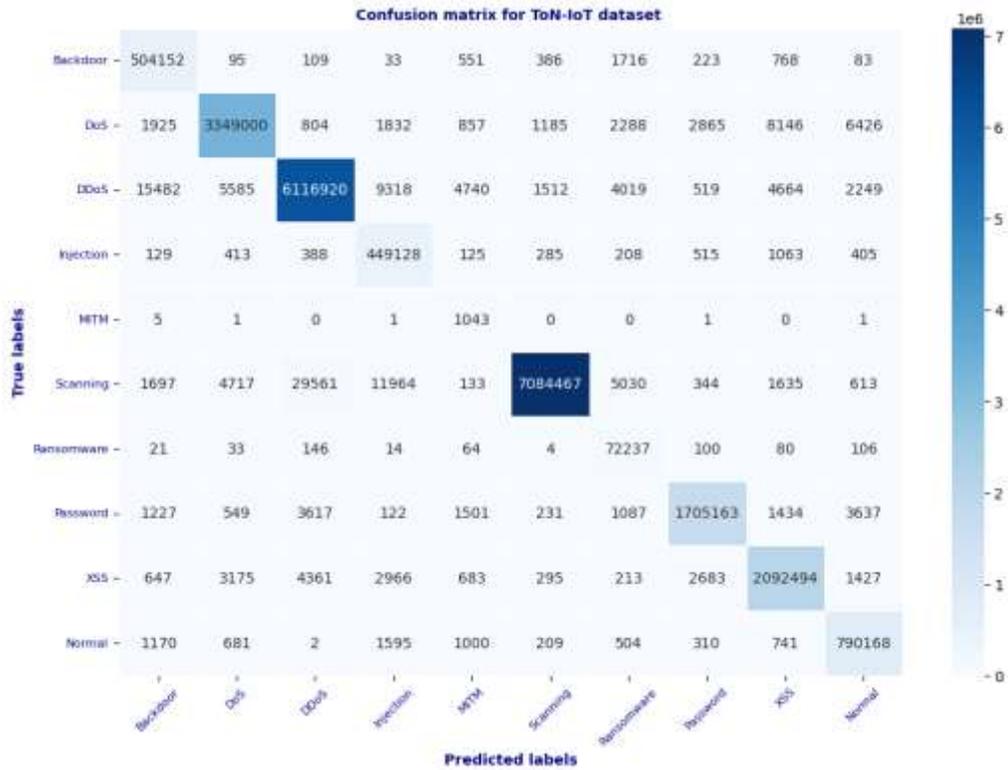

**Figure 7 .**Confusion matrix for the ToN-IoT dataset

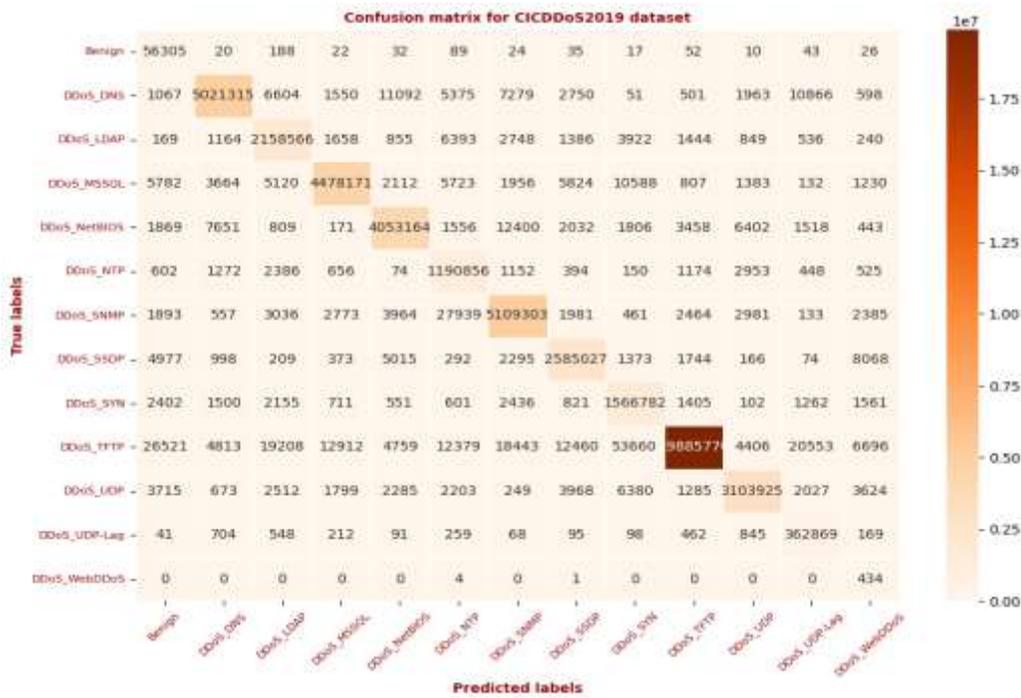

**Figure 8 .**Confusion matrix for the CICDDoS2019 dataset

Cluster 2 (1.5GHz CPU, 4GB RAM, 3 clients) also demonstrates strong performance with an accuracy of 98.88%, benefiting from its robust hardware that supports efficient data aggregation and model training through federated learning. Cluster 1, with the most modest specifications (1.2GHz CPU, 1GB RAM, 2 clients), still performs remarkably well with an accuracy of 98.68%, underscoring the effectiveness of the lightweight MobileNet architecture and the hybrid Gumbel-SoftMax loss function in achieving high performance even with limited resources.

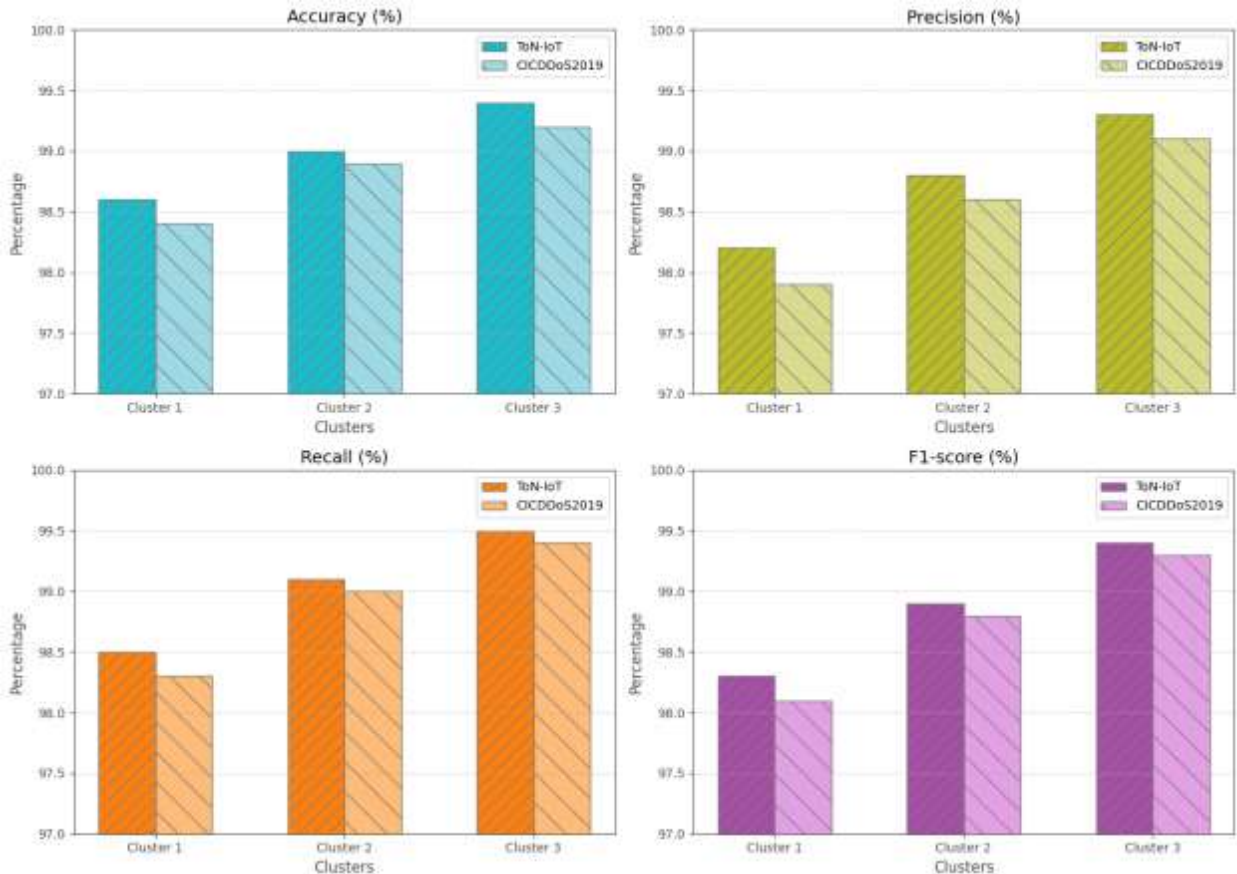

**Figure 9**. The performance metrics (Accuracy, Precision, Recall, F1-score) of the proposed method were evaluated for the ToN-IoT and CICDDOs2019 datasets, categorized into three clusters.

For CICDDoS2019, cluster 3 achieves superior results with an accuracy of 99.20%, precision of 99.35%, recall of 99.55%, and an F1-score of 99.40%. Cluster 2 also demonstrates strong performance with an accuracy of 98.88%, precision of 98.65%, recall of 99.15%, and an F1-score of 98.90%. Cluster 1 performs effectively with an accuracy of 98.45%, precision of 97.85%, recall of 98.25%, and an F1-score of 98.05%, highlighting the efficiency of Cluster-Based Federated Intrusion Detection with Lightweight Networks for Heterogeneous IoT method. Overall, Cluster 3, which has the highest computational resources, shows the best performance, achieving an accuracy on ToN-IoT and CICDDoS2019. In comparison, Cluster 1, with the least resources, demonstrates lower but still robust performance, particularly on the ToN-IoT dataset .However, CICDDoS2019's performance is slightly lower, reflecting the increased difficulty of detecting DDoS attacks. Notably, ToN-IoT generally outperforms CICDDoS2019 across all clusters. The CICDDoS2019 dataset includes Denial of Service (DoS) attacks, which are among the most challenging to detect due to their complexity and the significant traffic they generate. The emergence of heterogeneous IoT environments exacerbates this challenge, as varied device capabilities and network conditions make consistent detection difficult. Despite this, the proposed LightFedGSM-IDS method demonstrates robust performance across both datasets, showcasing its effectiveness in handling diverse and complex attack scenarios. The results highlight the model's adaptability and resilience, essential for securing modern IoT infrastructures. The proposed method leverages these lightweight network and new loss functions to ensure robust and reliable intrusion detection across all clusters, effectively addressing the challenges of resource limitations and enhancing model accuracy and generalization in heterogeneous IoT environments.

For evaluating latency, two scenarios are considered: traditional federated learning intrusion detection and the proposed method. Table 4 compares the latency performance between two approaches. The proposed method shows a marked improvement in performance compared to another approach. For ToN-IoT, firstly, the training time(end-to-end) has been significantly reduced from 2918 seconds to 1186 seconds which 2.47 times faster, thanks to the use of lightweight networks that lower computational complexity and resource consumption. This reduction is crucial as it allows the model to be trained more efficiently, even on devices with limited resources. Additionally, the testing latency has been decreased from 5.11 milliseconds to 2.36 milliseconds, enhancing the system's ability to provide real-time intrusion detection, which is essential in IoT environments where timely responses are critical. For CICDDOs2019 can be seen in Table 5. However, the lightweight network architecture plays a significant role in reducing latency. By utilizing a more efficient network structure, the proposed method minimizes the computational load and accelerates both the training and testing phases. Such efficiency is particularly beneficial in resource-scarce IoT networks, in which real-time performance and good quality intrusion detection require minimized latency. The time overhead for each round is acceptable as learning in the FL system is infrequent after convergence is attained. Consequently, this overhead can be safely overlooked in most instances. Moreover, the trade-off between total training time and time for each iteration is acceptable in consideration of different computational capacities and resources available in different IoT devices and being able to contribute different models for local malware identification efficiently. Overall, the method presented in this paper not only saves training time and latency but is also efficient in system performance improvement as a whole and is applicable for deployment in heterogeneous IoT networks.

**Table 4**. Latency comparison of Fed-IDS[] and proposed method (s: seconds, ms: milliseconds)

| Metric | Fed-IDS[] | | Proposed method | |
|---|---|---|---|---|
| | **ToN-IoT** | **CICDDOs2019** | **ToN-IoT** | **CICDDOs2019** |
| Training time ( end-to-end) | 2918 s | 2201 s | 1186 s | 981 s |
| Test latency | 5.11 ms | 4.01 ms | 2.36 ms | 1.85 ms |
| Time per round | 28.18 s | 21.52 s | 29.93 s | 24.29 |

### 4.6. Comparison between the proposed method and other methods

Table 5 presents a comprehensive comparative performance analysis for the novel federated intrusion detection method based on lightweight networks for heterogeneous IoT and state-of-the-art solutions. The performance is compared using two benchmarking datasets widely established in the research community: ToN-IoT and CICDDoS2019. The results exhibit better accuracy, precision, recall, and F1-score, thereby validating superiority for the novel method over existing solutions. On the ToN-IoT dataset, the proposed method achieves an outstanding accuracy of 99.09%, surpassing the performance of Kumar et al. [33] and Zainudin et al. [34]. This improvement highlights the effectiveness of the proposed approach in handling complex and heterogeneous IoT environments. Moreover, in terms of precision, recall, and F1-score, the proposed method records values of 98.91%, 98.76%, and 98.43%, respectively. These results indicate a substantial enhancement in classification performance, ensuring more accurate intrusion detection with fewer false positives and negatives. In the context of the CICDDoS2019 dataset, this method still excels compared to current methods in attaining a better level of accuracy at 99.22% than Ahmad et al. [19] and

Wang et al. [35]. Also, in aspects of precision (99.01%), recall (98.58%), and F1-score (99.16%), this method in this paper is better for detection. The improvement establishes generalizability and robustness of this method in different cases for IoT-based intrusion detection. The results from both benchmarks clearly indicate that the proposed method consistently outperforms existing state-of-the-art techniques. The enhanced accuracy, precision, recall, and F1-score highlight the efficiency and reliability of the cluster-based federated intrusion detection approach. By leveraging lightweight networks, the proposed method ensures optimized computational efficiency while maintaining high detection performance. This makes it a promising solution for real-world IoT security applications, offering improved protection against cyber threats and attacks in heterogeneous environments.

Table 5. Comparison of our model with existing state-of-the-art methods.

| Research | Dataset | | Accuracy | Precision | Recall | F1-score |
|---|---|---|---|---|---|---|
| | ToN-IoT | CICDDOs2019 | | | | |
| Kumar et al. [33] | | * | 83.5 | 73.5 | 70.01 | 66.7 |
| Zainudin et al. [34] | | * | 98.37 | - | - | - |
| **Proposed method** | | * | **99.09** | **98.91** | **98.76** | **98.43** |
| Wang et al. [35] | * | | 97.23 | 96.67 | 96.94 | 97.06 |
| Ahmad et al. [19] | * | | 98.85 | - | - | 99.02 |
| **Proposed method** | * | | **99.22** | **99.01** | **98.58** | **99.16** |

## 5. Conclusion & future work

The rapid expansion of heterogeneous Internet of Things (IoT) devices has introduced significant security challenges due to their vulnerability to sophisticated cyberattacks. To address these issues, we present the Cluster-Based Federated Intrusion Detection with Lightweight Networks for Heterogeneous IoT, a novel approach that enhances security while preserving data privacy and minimizing resource consumption. By clustering IoT devices based on hardware similarities, our approach facilitates efficient model training and aggregation tailored to each cluster's capabilities. The integration of the lightweight MobileNet model and a hybrid loss function combining Gumbel-SoftMax and SoftMax optimizes resource usage and significantly reduces latency while ensuring high detection accuracy. Evaluation using the ToN-IoT and CICDDoS2019 datasets reveals that the end-to-end training time is only 2.47 times higher than traditional federated learning, with a testing latency reduction of 2.16 times. Achieving accuracy rates of 99.22% and 99.02% for the ToN-IoT and CICDDoS2019 datasets, respectively, the proposed system demonstrates robust and reliable intrusion detection in diverse IoT environments. In summary, the cluster-based federated intrusion detection with lightweight networks marks a significant advancement in IoT security, effectively tackling the challenges of data heterogeneity and limited device capabilities while providing high-performance intrusion detection across various operational contexts. One the other hand, increasing the number of clusters significantly impacts the training time due to the exponential growth in model complexity. The system will require more computational power to update and aggregate model parameters across all clusters. Additionally, the communication overhead will increase, leading to higher latency. To mitigate this, optimizing the clustering mechanism and implementing asynchronous aggregation techniques can be effective in maintaining training efficiency.